\documentclass[iop,apj]{emulateapj}
\usepackage{graphicx}
\usepackage{verbatim}
\usepackage{parskip}
\usepackage{hyperref,soul}

\newcommand{\be}{\begin{itemize}}
\newcommand{\ee}{\end{itemize}}

\newlength\mystoreparindent

\def\fermi{\textit{Fermi }}

\received{September 18, 2017}
\accepted{November 16, 2017}
\journalinfo{Astrophysical~Journal~Supplement~Series}
\submitted{}
\shorttitle{MOJAVE. XV. VLBA Maps of Parsec-Scale AGN Jets}
\shortauthors{M. L. Lister et al.}

\begin{document}
\title{MOJAVE: XV. VLBA 15 GHz Total Intensity and Polarization Maps of 437 Parsec-Scale AGN Jets From 1996--2017}

\author{ M. L. Lister\altaffilmark{1},
M. F. Aller\altaffilmark{2},
H. D. Aller\altaffilmark{2},
M. A. Hodge\altaffilmark{1},
D. C. Homan\altaffilmark{3},
Y. Y. Kovalev\altaffilmark{4,5,6},
A. B. Pushkarev\altaffilmark{7,4},
T. Savolainen\altaffilmark{8,9,6} 
}

\altaffiltext{1}{
Department of Physics and Astronomy, Purdue University, 525 Northwestern Avenue,
West Lafayette, IN 47907, USA;
}
\altaffiltext{2}{
Department of Astronomy, University of Michigan, 311 West Hall, 1085
S. University Avenue, Ann Arbor, MI 48109, USA;
}

\altaffiltext{3}{
Department of Physics, Denison University, Granville, OH 43023, USA;}

\altaffiltext{4}{
Astro Space Center of Lebedev Physical Institute,
Profsoyuznaya 84/32, 117997 Moscow, Russia;
}
\altaffiltext{5}{Moscow Institute of Physics and Technology, Dolgoprudny, Institutsky per. 9, Moscow region, 141700, Russia}
\altaffiltext{6}{
Max-Planck-Institut f\"ur Radioastronomie, Auf dem H\"ugel 69,
53121 Bonn, Germany;
}

\altaffiltext{7}{
Crimean Astrophysical Observatory, 98409 Nauchny, Crimea, Russia;}

\altaffiltext{8}{ Aalto University Department of Electronics and Nanoengineering, PL 15500, FI-00076 Aalto, Finland}

\altaffiltext{9}{ Aalto University Mets\"ahovi Radio Observatory, Mets\"ahovintie 114, FI-02540 Kylm\"al\"a, Finland}



\begin{abstract}

  We present 5321 milliarcsecond-resolution total intensity and linear
  polarization maps of 437 active galactic nuclei (AGNs) obtained with
  the VLBA at 15 GHz as part of the MOJAVE survey, and also from the
  NRAO data archive. The former is a long-term program to study the
  structure and evolution of powerful parsec-scale outflows associated
  with AGNs.  The targeted AGNs are drawn from several flux-limited
  radio and $\gamma$-ray samples, and all have correlated VLBA flux
  densities greater than $\sim 50$ mJy at 15 GHz.  Approximately 80\%
  of these AGNs are associated with $\gamma$-ray sources detected by
  the \fermi LAT instrument.  The vast majority were observed with the
  VLBA on 5 to 15 occasions between 1996 January 19 and 2016 December
  26, at intervals ranging from a month to several years, with the
  most typical sampling interval being six months.  A detailed
  analysis of the linear and circular polarization evolution of these
  AGN jets are presented in other papers in this series.

\end{abstract}
 

\section{INTRODUCTION} 
\label{s:intro}

The collimated outflows from powerful active galactic nuclei (AGNs)
represent some of the most long-lived and energetic processes ever
discovered, and are now recognized to have played a prominent role in
early structure formation of our Universe via feedback mechanisms
(e.g., \citealt{2014ARAA..52..589H,2012ARAA..50..455F}). Due to their
exceedingly compact synchrotron radio emission, AGN jets have long
been a primary target of Very Long Baseline Interferometry (VLBI)
observations, which provide sub-milliarcsecond resolution maps on
scales out to several hundred parsecs from the central engine. With
the construction of the world's first dedicated multi-element VLBI
system, the Very Long Baseline Array (VLBA) in 1994, it became
possible to study large numbers of jets with full polarimetry at
regular intervals. The structure and evolution of linearly and
circularly-polarized jet emission provides crucial data for
understanding many poorly understood aspects of the flows, including
their collimation mechanisms, instability modes, magnetic field
structure, and the nature of the plasma itself (see, e.g.,
\citealt{2013EPJWC..6106001W,2015fers.confE..72A}). In 2002 we began a
follow-up program to the 2 cm VLBA survey of
\cite{1998AJ....115.1295K}, named MOJAVE (Monitoring of Jets in AGNs
with VLBA Experiments) with the goal of studying the parsec-scale
evolution of a large, complete sample of AGN jets with full
polarization VLBA imaging. The MOJAVE sample was subsequently expanded
to encompass an additional set of AGN jets detected in $\gamma$-rays
by the \fermi LAT instrument, and other compact radio-loud AGNs of
interest to the community. More details on individual objects and
their total intensity evolution  can be found in
other papers in this series (e.g., \citealt{MOJAVE_X, MOJAVE_XIII}).

In this paper we present multi-epoch 15 GHz VLBA total intensity and
linear polarization maps of 437 AGNs observed between 1996 January 19
and 2016 December 26, primarily as part of the MOJAVE program, with
supplementary data obtained from the NRAO archive. In \cite{MOJAVE_I}
we discussed the first epoch linear polarization maps of the original
MOJAVE flux-density limited AGN sample, and we described their
circular polarization properties in \cite{MOJAVE_II}.  We will present
a full multi-epoch analysis of the linear and circular polarization
characteristics of our extended sample of 437 AGNs in future papers.

\section{OBSERVATIONAL DATA}

\subsection{Sample Selection}
The VLBA maps in this paper are of 437 compact, radio-loud AGNs
observed as part of the MOJAVE program.  The latter originally started
as a complete set of 135 AGNs with J2000 declination $> -20\arcdeg$ and
galactic latitude $|b| > 2.5\arcdeg$ whose 15 GHz VLBA flux density
exceeded 1.5 Jy ($\ge 2$ Jy for sources below the celestial equator)
at any epoch between 1994.0 and 2004.0 \citep{MOJAVE_V}.  This
radio-selected sample was later expanded to encompass all 181 AGNs
above declination $-30\arcdeg$ known to have exceeded 1.5 Jy in VLBA
15 GHz flux density at any epoch between 1994.0 and 2010.0
\citep{MOJAVE_X}.

\begin{deluxetable}{lcc|c} 
\tablecolumns{4} 
\tablewidth{0pt}  
\tablecaption{\label{stattests} AGN Sample Properties}  
\tablehead{\colhead{Optical}  &
\colhead{\it{Fermi}} & \colhead{TeV} & \colhead{Total} \\
\colhead{Class} & 
\colhead{$\gamma$-ray} & \colhead{$\gamma$-ray} \\
\colhead{(1)} & \colhead {(2)}  &
\colhead{(3)} & \colhead{(4)} }
\startdata 
Quasar           & 203  & 5  & 265\\
BL Lac           & 121  &28  & 127\\
Radio Galaxy     &  10  & 2  &  27\\
Narrow Line Sy 1 &   5  & 0  &   5\\
Unidentified     &  12  & 0  &  13\\ 
\hline
Total            & 351  &35  & 437
\enddata
\tablecomments{Columns are as follows: (1) optical classification, (2) known associations with {\it Fermi} LAT $\gamma$-ray source, (3) TeV $\gamma$-ray sources in TEVCAT, (4) total number of sources.}
\end{deluxetable} 

With the launch of the {\it Fermi} observatory in 2008, two new
$\gamma$-ray-based AGN samples were added to MOJAVE.  The 1FM sample
\citep{2011ApJ...742...27L} consists of all 116 AGNs in the 1FGL
catalog \citep{1FGL} above declination $-30\arcdeg$ and galactic
latitude $|b| > 10\arcdeg$ with average integrated {\it Fermi} LAT $>0.1$
GeV energy flux above $3 \times 10^{-11}\, \mathrm {erg\,
  cm^{-2}\,s^{-1}}$.  In 2013, a new hard spectrum MOJAVE sample
of 132 AGNs was added, which had declination $> -20\arcdeg$, total
15 GHz VLBA flux density $> 0.1$ Jy, and mean {\it Fermi} 2LAC \citep{2LAC}
or 3LAC catalog \citep{3LAC} spectral index harder than 2.1. 

Of the 437 AGNs presented in this paper, 319 are members of one or
more of the above samples. An additional 105 AGNs are members of
either the pre-cursor survey to MOJAVE (the 2cm VLBA survey;
\citealt{1998AJ....115.1295K}), the low-luminosity MOJAVE AGN sample
\citep{MOJAVE_X}, the 3rd EGRET $\gamma$-ray catalog \citep{Hartman99},
the 3FGL {\it Fermi} LAT $\gamma$-ray catalog \citep{3FGL}, or the
ROBOPOL optical polarization monitoring sample
\citep{2014MNRAS.442.1693P}. Finally, we have included 13 AGNs that
were originally candidates for the above samples, but did not meet the
final selection criteria.

A summary of the sample in terms of optical classification and
$\gamma$-ray emission can be found in Table 1. The common feature of
these AGNs is that they all have bright, compact radio emission on
milliarcsecond scales ($\gtrsim 50$ mJy at 15 GHz). Since such
emission is usually associated with relativistically beamed jets, the
sample is primarily composed of flat-spectrum radio quasars (61\%) and
BL Lac objects (29\%), which have powerful jets oriented at small
angles to the line of sight (i.e., blazars). A total of 27 radio
galaxies have made it into the sample due to their low redshift, or as
a result of having GHz-peaked radio spectra. Additionally, there are 5
narrow line Seyfert I galaxies, and 14 AGN which lack identified
optical counterparts. A large fraction of the sample (80\%) are
associated with 0.1 GeV -- 100 GeV $\gamma$-ray detections made by the
{\it Fermi} LAT instrument, and 35 AGNs are listed as known TeV
$\gamma$-ray emitters in TEVCAT\footnote{http://tevcat.uchicago.edu}.

\subsection{VLBA Observations}

The data consist of 5321 observations of 437 AGNs in the 15 GHz band,
obtained between 1996 January 19 and 2016 December 26 with the VLBA in
full polarization mode (Table~\ref{maptable}).  All targets were
bright enough for direct interferometric fringe detection with 2048
Mbps VLBA observations at 15 GHz (i.e., $\gtrsim 50$ mJy) during at
least one epoch.  In the case of two partially resolved AGNs
(TXS~0831$+$557 and TXS~0954$+$556), no interferometric fringes were
found at any epoch on the longest (Mauna Kea and St. Croix) baselines
during data processing.  

\begin{deluxetable*}{llllcrrcccccrc} 
\tablecolumns{14} 
\tabletypesize{\scriptsize} 
\tablewidth{0pt}  
\tablecaption{\label{maptable}Summary of 15 GHz Image Parameters}  
\tablehead{ & &  & 
\colhead{$B_\mathrm{maj}$} &\colhead{$B_\mathrm{min}$} & \colhead{$B_\mathrm{pa}$} &  
\colhead{$I_\mathrm{tot}$} &  \colhead{$\sigma_\mathrm{I}$}  &  \colhead{$I_\mathrm{base}$} & 
\colhead{$P_\mathrm{tot}$} & \colhead{$\sigma_\mathrm{Q,U}$}  & \colhead{$P_\mathrm{base}$}&\colhead{EVPA} &  \colhead{Fig.} \\ 
\colhead{Source} & \colhead{Alias} & \colhead {Epoch} & 
\colhead{(mas)} &\colhead{(mas)} & \colhead{(\arcdeg)} &  
\colhead{(mJy)} & \colhead{(mJy} & \colhead{(mJy}  &  \colhead{(mJy) } &   \colhead{(mJy}  &  \colhead{(mJy }& \colhead{(\arcdeg)} & \colhead{Num.} \\ 
 &  &  &   & 
 & &   
 & \colhead{beam$^{-1}$)}  &  \colhead{beam$^{-1}$)} & & \colhead{beam$^{-1}$)}  &  \colhead{beam$^{-1}$)} & & \\ 
\colhead{(1)} & \colhead{(2)} & \colhead{(3)} & \colhead{(4)} &  
\colhead{(5)} & \colhead{(6)} & \colhead{(7)} & \colhead{(8)} & \colhead{(9)}& \colhead{(10)} &  \colhead{(11)} & \colhead{(12)} &   \colhead{(13)} &  \colhead{(14)} } 
\startdata 
0003+380 & S4 0003+38  & 2006 Mar 9 & 1.01 & 0.73 & 18 & 650 & 0.44 & 1.15 & 16 & 0.45 & 1.48 & $78$  & 4.1   \\ 
 &   & 2006 Dec 1 & 0.85 & 0.58 & $-$17 & 512 & 0.41 & 1.10 & 9.2 & 0.46 & 1.42 & $146$  & 4.2   \\ 
 &   & 2007 Mar 28 & 0.86 & 0.61 & $-$15 & 604 & 0.33 & 0.80 & 5.6 & 0.35 & 1.08 & $144$  & 4.3   \\ 
 &   & 2007 Aug 24 & 0.92 & 0.58 & $-$28 & 555 & 0.25 & 0.80 & 7.3 & 0.30 & 0.88 & $166$  & 4.4   \\ 
 &   & 2008 May 1 & 0.82 & 0.57 & $-$9 & 808 & 0.24 & 0.70 & 12 & 0.26 & 0.84 & $178$  & 4.5   \\ 
 &   & 2008 Jul 17 & 0.84 & 0.55 & $-$12 & 727 & 0.22 & 0.60 & 17 & 0.25 & 0.79 & $175$  & 4.6   \\ 
 &   & 2009 Mar 25 & 0.84 & 0.62 & $-$12 & 435 & 0.16 & 0.50 & 10 & 0.17 & 0.56 & $145$  & 4.7   \\ 
 &   & 2010 Jul 12 & 0.89 & 0.54 & $-$12 & 438 & 0.17 & 0.50 & 10 & 0.19 & 0.63 & $132$  & 4.8   \\ 
 &   & 2011 Jun 6 & 0.91 & 0.54 & $-$10 & 605 & 0.18 & 0.50 & 2.1 & 0.20 & 0.63 & $108$  & 4.9   \\ 
 &   & 2013 Aug 12 & 0.84 & 0.53 & $-$4 & 668 & 0.20 & 0.50 & 9.8 & 0.21 & 0.75 & $139$  & 4.10   \\ 
0003$-$066 & NRAO 005  & 2003 Feb 5 & 1.32 & 0.53 & $-$6 & 2842 & 0.31 & 1.40 & 103 & 0.37 & 1.00 & $17$  & 4.11   \\ 
 &   & 2004 Jun 11 & 1.30 & 0.49 & $-$7 & 3273 & 0.32 & 1.40 & 255 & 0.61 & 1.57 & $22$  & 4.12   \\ 
 &   & 2005 Mar 23 & 1.32 & 0.53 & $-$5 & 3034 & 0.27 & 1.50 & 129 & 0.32 & 1.38 & $4$  & 4.13   \\ 
 &   & 2005 Jun 3 & 1.37 & 0.53 & $-$8 & 3017 & 0.23 & 1.40 & 110 & 0.30 & 1.00 & $175$  & 4.14   \\ 
\enddata 
\tablecomments{The full table is available electronically from arXiv and the journal.
Columns are as follows: (1) B1950 name, (2) other name, (3) date of VLBA observation, (4) FWHM major axis of restoring beam (milliarcseconds), (5) FWHM minor axis of restoring beam (milliarcseconds), (6) position angle of major axis of restoring beam (degrees), (7) total cleaned I flux density (mJy),  (8) rms noise level of Stokes I image (mJy per beam), (9) lowest I contour level (mJy per beam), (10) total cleaned P flux density (mJy), or upper limit, based on 3 times the P rms noise level,  (11) average of blank sky rms noise level in Stokes Q and U images (mJy per beam), (12) lowest linear polarization contour level (mJy per beam), (13) integrated electric vector position angle (degrees), (14) figure number.}
\tablenotetext{a}{NRAO archive epoch}
\end{deluxetable*} 

Most of the VLBA observations (86\%) were carried out as part of the
MOJAVE program \citep{MOJAVE_V}, while the remainder were downloaded
from the NRAO archive\footnote{http://archive.nrao.edu} and processed
for the purpose of increasing the number of epochs on particular AGNs.
Our minimal criteria for the individual observations were: (i) at
least 8 of 10 VLBA antennas present, (ii) either the Mauna Kea or the
St. Croix antenna present for some portion of the observation, (iii) a
minimum of 3 scans reasonably separated over a range of hour angle,
(iv) available scans on suitable electric vector position angle and
instrumental polarization (feed leakage) calibrators during the
observing session.

\begin{figure}[t]
\centering
\includegraphics[trim=0cm 0cm 0cm 0cm,clip,width=0.47\textwidth]{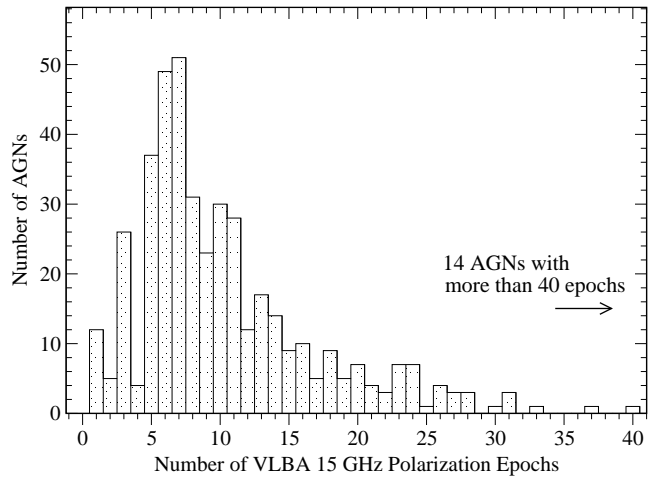}
\caption{\label{nepoch_histogram} 
Distribution of number
of 15 GHz VLBA polarization observations for 437 AGNs in the
MOJAVE program.  }
\end{figure}

\begin{figure}
\centering
\includegraphics[trim=0cm 0cm 0cm 0cm,clip,width=0.47\textwidth]{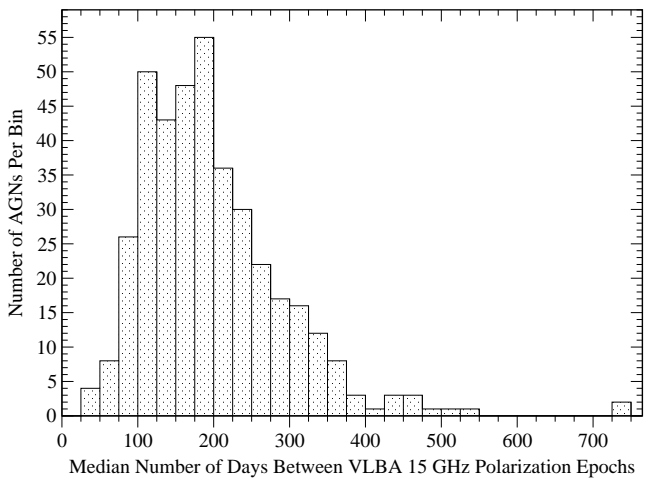}
\caption{
\label{epochinterval_histogram} 
Distribution of
median interval between 15 GHz VLBA polarization observations for
437 AGNs in the MOJAVE program.
}
\end{figure}

Most of the AGNs have between 5 and 15 VLBA observation epochs,
although a substantial number have considerably more epochs (up to
128, for BL Lac=TXS~2200$+$420), and 32 AGNs have four epochs or fewer
(Fig.~\ref{nepoch_histogram}). The median time interval between
observations varies from 35 days for BL Lac to roughly two
years for TXS~2021$+$614 and PKS~B1345$+$125
(Fig.~\ref{epochinterval_histogram}). This large range reflects the
observational strategy of the MOJAVE program, which aims to
observe each AGN jet with a cadence appropriate for its angular
expansion speed \citep{MOJAVE_XIII}.

We calibrated and edited the data in AIPS using the standard
techniques described in the AIPS Cookbook, and exported them to the
DIFMAP package \citep{DIFMAP} for imaging. A full description of the
method can be found in \cite{MOJAVE_V}.  We determined the amplitudes
and phases of the complex feed leakage terms for each IF and antenna
by taking the median of the solutions obtained from the AIPS task
LPCAL on bright, core-dominated AGNs that were either unpolarized or
had relatively simple polarization structures.  For the majority of
observations obtained prior to 2013 September 1, we determined the
instrumental electric vector position angle (EVPA = (1/2) arctan
($U/Q$), where $Q$ and $U$ are Stokes parameters) rotation by using
several leakage term phases that were constant over time at certain
antennas (see, e.g., \citealt{Gomez2002}). This method requires an
external calibration to establish the absolute EVPA directions on the
sky; for this purpose we used near-simultaneous single-dish
measurements of selected core-dominated jets at 14.5 GHz made at the
U. Michigan Radio Observatory, up until its closure in 2012 July. With
the introduction of new digital electronics at the VLBA antennas in
mid-2013, the leakage term phases became unstable over time.  For
observations after this date we therefore relied on optically thin jet
features in multiple AGNs that we have found to have relatively stable
EVPAs for calibration purposes.  We also used this method for several
NRAO archive observations that used a different recording setup from
our regular MOJAVE observations, and thus had different leakage term
phases.  Based on our comparisons of highly compact AGNs to
near-simultaneous single-dish observations, we estimate our VLBA EVPA
measurements are accurate to $\sim 5\arcdeg$, and our flux densities
to $\sim 5$\%.

\subsection{Map Characteristics}

\begin{figure}
\centering
\includegraphics[trim=0cm 0cm 0cm 0cm,clip,angle=-90,width=0.47\textwidth]{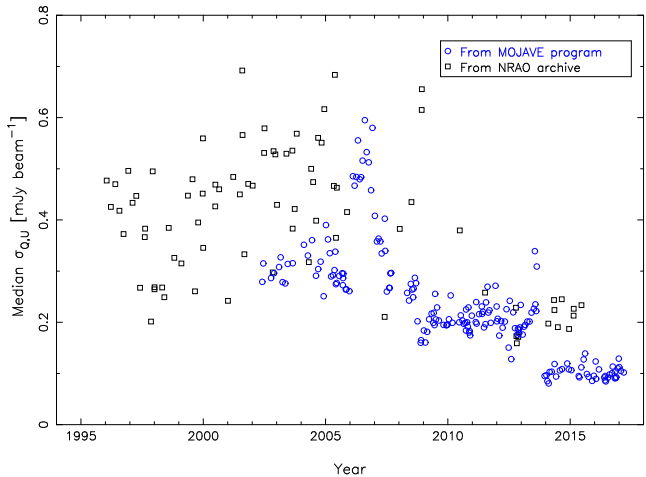}
\caption{\label{obscode_qurms} 
Plot of median Stokes Q,U blank-sky r.m.s. noise level for individual
observing sessions from the MOJAVE program (blue circles) and the NRAO
archive (black crosses). The increased noise in MOJAVE maps in 2006--2007 is due to the addition of other VLBA observing frequencies during that period (see \citealt{MOJAVE_VIII}).
}
\end{figure}

We measured the rms noise levels ($\sigma$) in the Stokes I, Q, and U
maps by shifting the map center by 1 arcsecond and calculating the
standard deviation of the (blank sky) pixel intensity distribution.
The Q and U maps typically had similar noise characteristics, so we
will hereafter refer to $\sigma_{\mathrm Q,U}$ as the average of these
two noise levels.  Our large set of maps has a factor of
$\sim$ 5 range of rms map noise level, due to differences in the total
integration time, number of scans and antennas present in each
observing session, and in the observed bandwidth (as set by the
maximum recordable bit rate). The latter has steadily increased over
time, due to hardware upgrades at the VLBA antennas.  This is apparent
in Figure~\ref{obscode_qurms}, where we plot the median $\sigma_{\mathrm
  Q,U}$ of the maps from each observing session having more than 4 AGNs
versus the date of observation. The most recent MOJAVE observations,
recorded at 2048 Mbps with 2 bit sampling for a total observing
bandwidth of 256 MHz per polarization, have typical map rms noise
levels of 0.1 mJy beam$^{-1}$ in Stokes I, Q and U.  This represents a
factor of $\sim 3$ improvement since the start of the MOJAVE VLBA
program in 2002.



As discussed by \cite{1974ApJ...194..249W} and
\cite{1985AA...142..100S}, the noise statistics of a linear
polarization map ($P = \sqrt{Q^2 + U^2}$) are Ricean, since it is a
vector addition of the Q and U maps, each of which have Gaussian noise
statistics. In the case of a blank sky region, or one with low
polarization signal to noise ratio, the noise statistics of the P map
approximately follow a Rayleigh distribution, which has a mean
$(\pi/2)^{1/2} \sigma_\mathrm{Q,U}$ and variance $(2 -
\pi/2)\sigma^2_\mathrm{Q,U}$.  It is only in high signal-to-noise
regions ($\gtrsim 3$) that the P map noise statistics approach a
Gaussian distribution, with variance $\sigma^2_\mathrm{Q,U}$. The
corresponding noise in the fractional polarization map will be
$\sigma_m = \sigma_{Q,U} / I$ (see, e.g., Appendix B of \citealt{MOJAVE_VIII}).

Additional complications arise with polarization imaging due to errors
associated with the antenna feed leakage terms
\citep{1994ApJ...427..718R}, and errors during the CLEAN process due
to incomplete (u,v) plane coverage \citep{MOJAVE_VIII}. Imperfect
correction of the leakage of polarization signal orthogonal to the
nominal polarization causes polarized emission to be spread throughout
the map in a non-uniform manner. Since these errors are proportional
to the total Stokes I intensity, they can be significant in maps of
bright radio sources.  This frequently leads to spurious high
fractional polarization values ($\gtrsim 40\%$) at the edges of the
jets in our maps, where the total intensity falls off exponentially.
These regions of the polarization maps should therefore be interpreted
with caution.

\subsection{Contour Maps}

\begin{figure}[t]
\centering
\includegraphics[trim=0cm 0cm 0cm 0cm,angle=270,clip,width=0.47\textwidth]{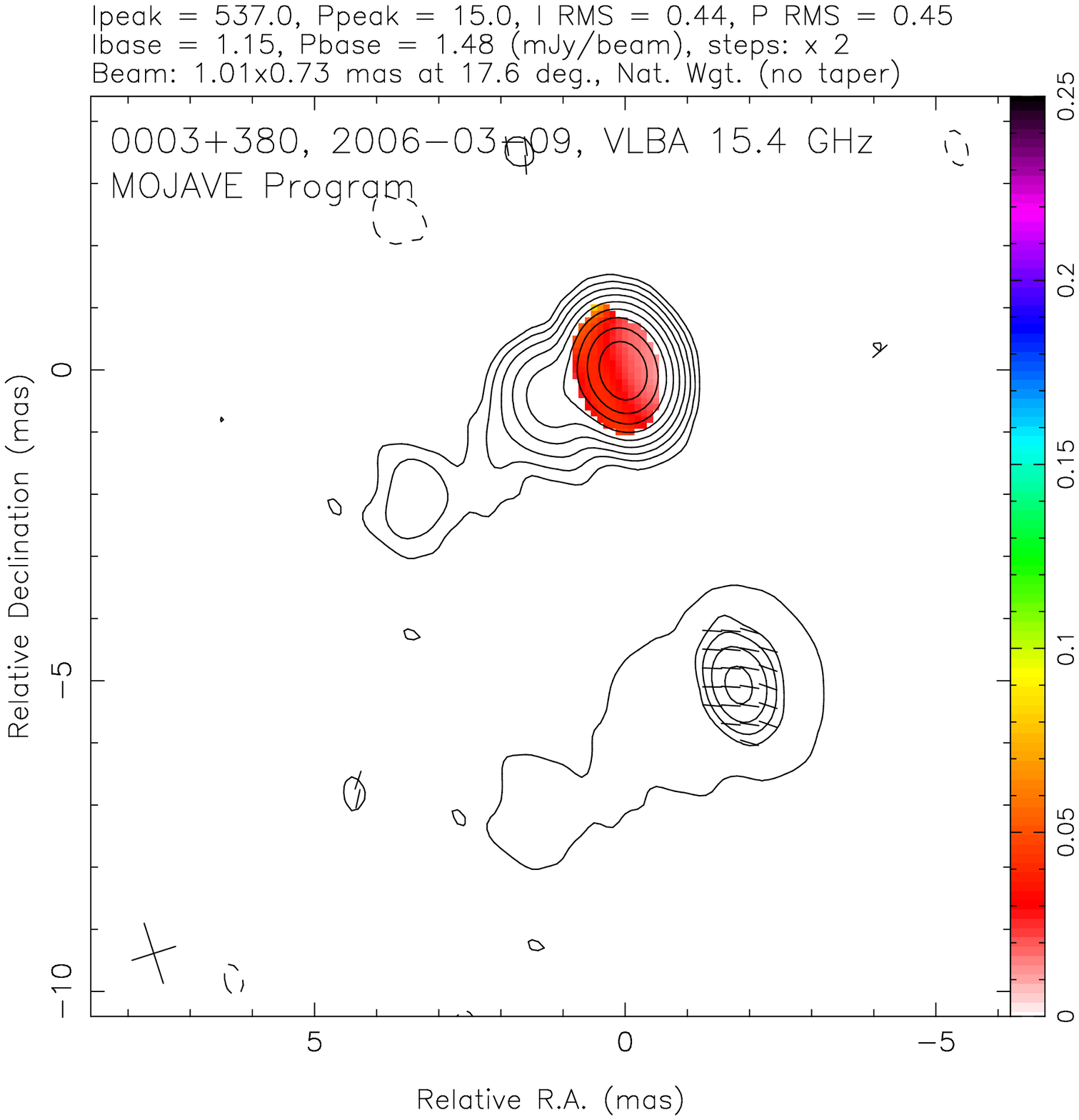}
\caption{\label{natwgtmaps} 
15 GHz VLBA maps of the MOJAVE AGN sample\footnote{\scriptsize{}\url{http://www.physics.purdue.edu/MOJAVE/MOJAVE_XV/natwgtmaps.tar}}.
Each panel contains two contour maps of the radio
  source, the first consisting of I contours in successive integer
  powers of two times the lowest contour level, with linear fractional
  polarization overlaid according to the color wedge. A single
  negative I contour equal to the base contour level is also plotted.
  The second map includes the lowest positive I contour from the
  first map, and linearly polarized intensity contours, also in
  increasing powers of two. The sticks indicate the electric
  polarization vector directions, uncorrected for Faraday rotation.
  The FWHM dimensions and orientation of the elliptical Gaussian
  restoring beam are indicted by a cross in the lower left corner of
  the map.}
\end{figure}

In Figure~\ref{natwgtmaps}, we present ``dual-plot'' maps of each VLBA
AGN observation at all available epochs. We previously published
first-epoch maps of this type for the original MOJAVE sample in
\cite{MOJAVE_I}. The FWHM dimensions and orientation of the
naturally-weighted elliptical Gaussian restoring beam are indicated by
a cross in the lower left corner of each map. The beam size varies
somewhat with the declination of the AGN and number of antennas, but
has typical dimensions of 1.1 mas $\times$ 0.5 mas. We gridded each of
the Stokes maps with a scale of 0.1 mas per pixel. We list the
parameters of the restoring beam, base contour levels, total cleaned
flux densities, and blank sky map noise levels for each map in
Table~\ref{maptable}.  The false color corresponds to fractional
polarization, and is superimposed on a total intensity contour map of
the radio emission.  The false-color scale is truncated at a
fractional polarization level of 50\%, since such high values are
generally not seen at the same locations in successive epochs on a
given jet, and are most likely spurious.  We also plot no fractional
polarization in regions that lie below the lowest total intensity
contour.  The latter typically corresponds to 3 times the rms noise
level of the map, although this can be higher in the cases of AGNs
with poorer interferometric coverage (due to very low or
near-equatorial declination) or very bright jet cores (due to dynamic
range limitations on the map).  Similarly, the lowest polarization
contour is typically 3 times $\sigma_\mathrm{Q,U}$, but in $\sim 15\%$
of the epochs is $>5$ times $\sigma_\mathrm{Q,U}$ due to a high peak
total intensity in the map, or polarization feed leakage errors. We
have not made any Ricean de-biasing corrections (i.e.,
$P_\mathrm{corr} = P_\mathrm{obs}\sqrt{1 - (\sigma_P /
  P_\mathrm{obs}^2})$ ) to the maps, since these are $\lesssim$ 5\%
for regions above our lowest polarization contour level.

Since any absolute sky positional information is lost due to
self-calibration, the map origin is placed at the total intensity Gaussian
model-fit position of the core feature, as described by
\cite{MOJAVE_XIII}. The latter is typically the brightest feature in
the map, located at the optically thick surface close to the base of
the jet. The positional accuracy of the core model-fit is typically 20\% of
the FWHM beam dimension listed in Table~\ref{maptable}, as discussed
in \cite{MOJAVE_VI}.

A second map of the jet emission is located at an arbitrary positional
offset from the first, and consists of polarized intensity contours,
and the lowest total intensity contour for reference.  The overlaid
sticks indicate the observed electric vector directions, and are of an
arbitrary fixed length.  We have not corrected their orientations for
any Faraday rotation either internal or external to the AGN jet.  Our
rotation measure study of the MOJAVE sample \citep{MOJAVE_VIII} showed
that the emission from most of these jets experiences only a few
degrees of Faraday rotation (typically in the region near the base of
the jet) at 15 GHz.  The total intensity and polarization contours are
plotted at increasing powers of 2 times the base contour level listed
on the map and in Table~\ref{maptable}.

\section{Time-lapse Movies}

\begin{figure}[t]
\centering
\includegraphics[trim=2cm 0cm 2cm 2cm,clip,width=0.5\textwidth]{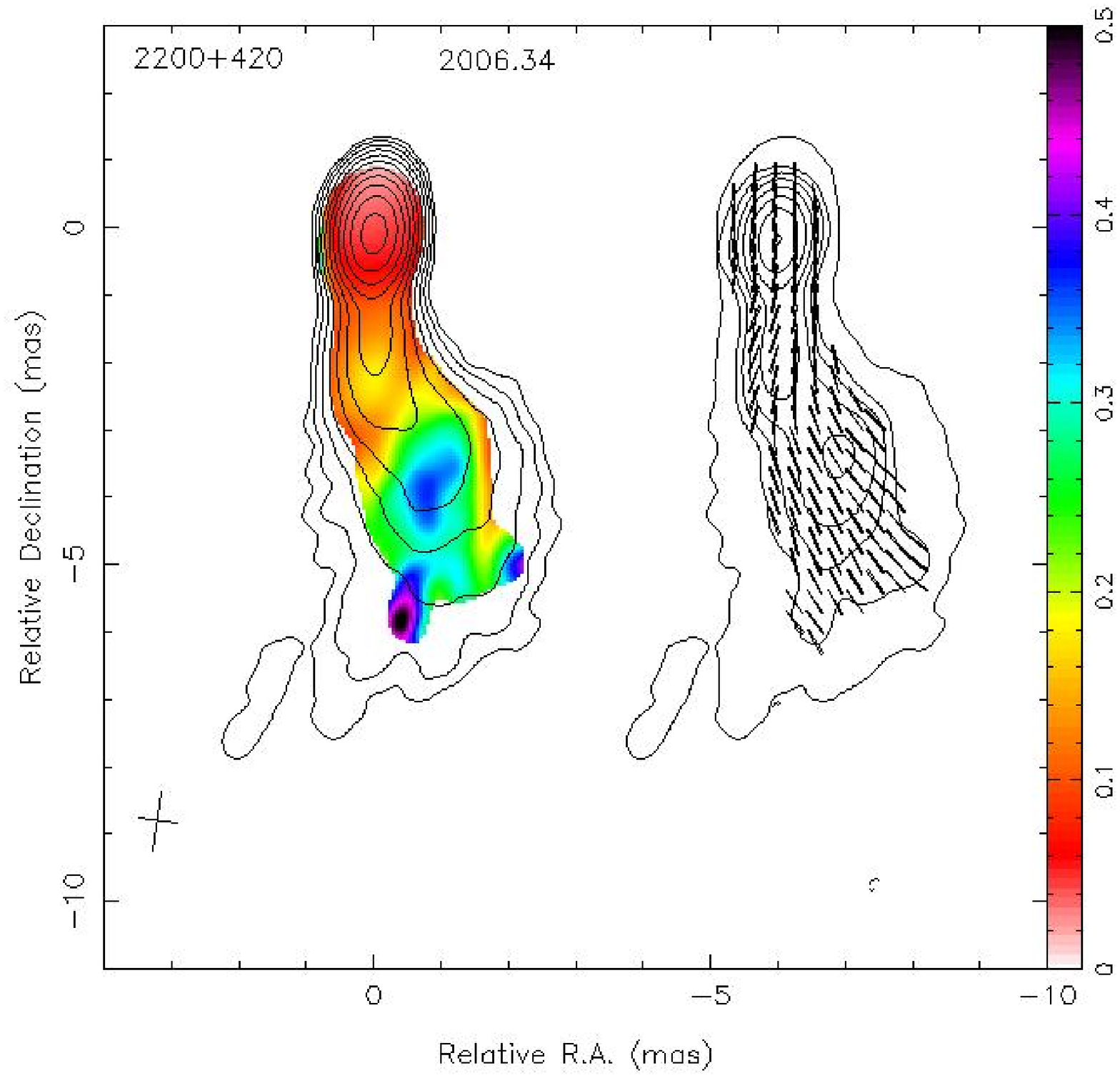}
\caption{
\label{movie} 
Linearly interpolated time-lapse
movie made from the multi-epoch VLBA maps for BL Lac\footnote{\scriptsize{}\url{http://www.physics.purdue.edu/MOJAVE/MOJAVE_XV/2200+420.pol.mpg}}.
    Each epoch
    was restored with a median beam, whose FWHM dimensions (0.87 mas
    $\times$ 0.56 mas at $-8.4\arcdeg$) are indicated by the cross in
    the lower left corner. The false color and contour scheme follows
    that use in Figure~\ref{natwgtmaps}. The base contours in total
    intensity and polarization are 2 mJy per beam, and one year of
    calendar time corresponds to 7 seconds in the time-lapse movie.
    }
\end{figure}

In Figure~\ref{movie} we show a time-lapse MPEG movie of the
multi-epoch VLBA polarization and total intensity maps for BL Lac from
1999 May 16 to 2016 Dec 26. Movies for other AGNs can be found on the
MOJAVE website\footnote{http://www.physics.purdue.edu/MOJAVE}.  We
constructed the movie using a two-point linear interpolation across
each successive epoch, treating each map pixel independently.  One
year of calendar time corresponds to 7 seconds in the time-lapse.
Prior to interpolation, we registered all of the epoch maps to the
Gaussian-fit core positions and restored them to a scale of 0.05
milliarcseconds per pixel using identical median beam dimensions (0.87
mas $\times$ 0.56 mas at $-8.4\arcdeg$) that were based on the full
set of naturally weighted VLBA epochs available for this AGN. The
individual frames of the movie follow the same format as the
individual epoch maps of Figure~\ref{natwgtmaps}, however the base
contours in total intensity and polarization are set to 2 mJy per beam
in all of the movie frames.

\section{Summary}

We have presented 5321 milliarcsecond-resolution, VLBA 15 GHz total
intensity and linear polarization maps of 437 AGN jets obtained
between 1996 Jan 19 and 2016 Dec 26 as part of the MOJAVE program, and
also from the NRAO data archive. These AGNs are drawn from several
flux-limited radio and $\gamma$-ray samples, and are all compact, with
correlated flux densities greater than $\sim 50$ mJy. Most were
observed on at least 5 to 15 occasions between 1996 and 2017, at
intervals ranging from a month to several years, with the most typical
interval being 6 months.  We have analyzed the multi-epoch total
intensity evolution of most of these jets in a series of papers
\citep{MOJAVE_VI, MOJAVE_VII, MOJAVE_X, MOJAVE_XII}, with the most
recent analysis presented by \cite{MOJAVE_XIII}. We will present a
detailed analysis of the linear and circular polarization evolution of
these AGN jets in upcoming papers in this series.

\acknowledgments

The MOJAVE project was supported by NASA-{\it Fermi} grants
NNX08AV67G, NNX12A087G, and NNX15AU76G. MFA was supported in part by
NASA-{\it Fermi} GI grants NNX09AU16G, NNX10AP16G, NNX11AO13G,
NNX13AP18G and NSF grant AST-0607523.  YYK and ABP were supported by
the Russian Foundation for Basic Research (project 17-02-00197), the
Basic Research Program P-7 of the Presidium of the Russian Academy of
Sciences and the government of the Russian Federation (agreement
05.Y09.21.0018).  TS was supported by the Academy of Finland projects
274477 and 284495. The Long Baseline Observatory and the National
Radio Astronomy Observatory are facilities of the National Science
Foundation operated under cooperative agreement by Associated
Universities, Inc.  This work made use of the Swinburne University of
Technology software correlator \citep{2011PASP..123..275D}, developed
as part of the Australian Major National Research Facilities Programme
and operated under licence.

\bibliographystyle{apj}
\bibliography{lister}

\end{document}